\documentclass[aps,prl,twocolumn,showpacs,preprintnumbers]{revtex4}
\usepackage{amsmath}
\usepackage{dcolumn}
\usepackage{bm}
\usepackage{subfigure}
\usepackage{amsfonts}
\usepackage{amssymb}
\usepackage{makeidx}
\usepackage{epsfig}
\usepackage{graphicx}

\begin{document}

\title{\textbf{Kinetic equilibria of relativistic collisionless plasmas\\
in the presence of non-stationary electromagnetic fields}}
\author{Claudio Cremaschini\thanks{%
Electronic-mail: claudiocremaschini@gmail.com}$^{a}$, Massimo Tessarotto$%
^{b,a}$ and Zden\v{e}k Stuchl\'{\i}k$^{a}$}
\affiliation{$^{a}$Institute of Physics, Faculty of Philosophy and Science, Silesian
University in Opava, Bezru\v{c}ovo n\'{a}m.13, CZ-74601 Opava, Czech Republic%
\\
$^{b}$Department of Mathematics and Geosciences, University of Trieste, Via
Valerio 12, 34127 Trieste, Italy}
\date{\today }

\begin{abstract}
The kinetic description of relativistic plasmas in the presence of
time-varying and spatially non-uniform electromagnetic fields is a
fundamental theoretical issue both in astrophysics and plasma physics. This
refers, in particular, to the treatment of collisionless and
strongly-magnetized plasmas in the presence of intense radiation sources. In
this paper the problem is investigated in the framework of a covariant
gyrokinetic treatment for Vlasov-Maxwell equilibria. The existence of a new
class of kinetic equilibria is pointed out, which occur for
spatially-symmetric systems. These equilibria are shown to exist in the
presence of non-uniform background EM fields and curved space-time. In the
non-relativistic limit this feature permits the determination of kinetic
equilibria even for plasmas in which particle energy is not conserved due to
the occurrence of explicitly time-dependent EM fields. Finally, absolute
stability criteria are established which apply in the case of infinitesimal
symmetric perturbations that can be either externally or internally produced.
\end{abstract}

\maketitle

\bigskip


\section{Introduction}

This paper deals with two critical aspects of relativistic theoretical
astrophysics and plasma kinetic theory. The first one concerns the
formulation of a non-perturbative covariant gyrokinetic theory for the
description of relativistic single-particle dynamics in curved space-time
and non-uniform electromagnetic fields. The gyrokinetic theory developed
here applies in the presence of both time- and space-varying electromagnetic
fields, e.g., due to radiation sources, and improves previous literature
treatments with the implementation of extended phase-space guiding-center
transformations of particle position and velocity 4-vectors. It is shown
that the single-particle magnetic moment associated with the Larmor rotation
of single charges around magnetic field lines is conserved as an integral of
motion under suitable assumptions and can be also determined as an adiabatic
invariant in principle with arbitrary accuracy. The second issue addressed
is related to the proof of existence of relativistic kinetic equilibria for
multi-species collisionless and magnetized plasmas in spatially-symmetric
configurations subject to non-stationary electromagnetic fields. It is shown
that equilibria of this type can be realized in terms of Gaussian-like
distributions as a consequence of the conservation of the particle magnetic
moment, and they are absolutely stable with respect to infinitesimal
axisymmetric perturbations.

For a start it is worth setting these topics in the proper physical
perspective. Indeed, the theory of kinetic equilibria in collisionless
magnetized plasmas presents in many aspects a formidable theoretical
challenge still to be tackled.

An occurrence of this type si related to the experimental evidence, both in
laboratory and astrophysical plasmas, of kinetic plasma regimes which
persist for long times (with respect to the observer and/or plasma
characteristic times), despite the presence of macroscopic time-varying
phenomena due to flows, non-uniform gravitational/EM fields and EM
radiation, such as that arising from single-particle radiation-reaction
processes \cite{maha-1,CR2011-RR}. The conjecture is that, for collisionless
plasmas, these states might actually correspond - at least locally and in a
suitable asymptotic sense - to some kind of kinetic equilibrium which
characterizes the species kinetic distribution function (KDF). This is
realized when the species KDFs are all assumed to be functions only of the
single particle exact and/or adiabatic invariants. It must be noticed that,
in the framework of a covariant description, equilibrium solutions of the
Vlasov equation can correspond to time-varying configurations relative to a
coordinate time, for example when referred to a non-relativistic inertial
reference frame, as can be the observer laboratory frame. As a consequence
of this definition, it follows that the concept of relativistic kinetic
equilibrium is not at variance with the presence of time-varying phenomena.

In this regard, a relevant issue concerns the possible effect of
non-uniform, i.e., both space and time varying, electromagnetic (EM) fields
acting on magnetized plasmas in the presence of some kind of spatial
symmetry. This refers in particular to possible explicit time dependences
arising in plasmas which\ can be treated as non-relativistic and are endowed
with a characteristic time-scale $\Delta t$ much larger than the Larmor time 
$\tau _{L}$ when observed in a suitable non-relativistic reference system.\
In\ particular, in the case of non-relativistic laboratory plasmas this may
be identified with the laboratory inertial frame. Instead, for astrophysical
plasmas that are characterized by non-relativistic temperature, a convenient
alternative choice may be provided by the fluid co-moving frame, which is
locally at rest with respect to the fluid element associated with the
plasma\ (the plasma may still be relativistic when seen from an observer's
inertial frame). In both cases $\Delta t$ can be defined as $\Delta t$ $%
=L/v_{th}$, where $L$ is the characteristic scale-length associated with the
fluid fields of the plasma, while $v_{th}=\sqrt{2T/M}$ is the ion thermal
velocity for an ion-electron plasma, with $T$ and $M$ being respectively the
ion temperature and mass.

However, both for laboratory and astrophysical plasmas, more general plasma
regimes can in principle occur, in which the charged particles of the plasma
are characterized by relativistic velocities. Configurations of this type
can arise, for instance, when intense external radiation sources are present
in strongly magnetized plasmas \cite{as7,as8,as2,as3,as1}. In astrophysics,
an epitome example is provided by the complex phenomenology associated with
accretion disc plasmas around compact objects, possibly associated with the
simultaneous occurrence of relativistic jets, which may be characterized by
the presence of time-varying EM fields, curved space-time as well as plasma
flows \cite{as9,as10,as6,as5,as4,stu99}. In the literature, achievements
concerning the theoretical investigations of both equilibrium and stability
properties of relativistic plasmas of this type have been obtained based on
fluid approaches. Examples are provided by Refs.\cite%
{ma-1,ma-2,ma-3,ma-4,pprd1,pprd2,pprd3,pprd4}. However, a theoretical
treatment of these phenomena in the context of kinetic theory remains
unsatisfactory to date, because of the difficulty of identifying the
appropriate kinetic regimes.\ In fact, at the microscopic level both
single-particle energy and magnetic moment might become non-conserved
dynamical variables. In such a case\textbf{\ }it is not known whether the
theory of single-particle dynamics based on gyrokinetic theory (GKT) \cite%
{Little83} and in particular its extension in the presence of flows \cite%
{Catto1987,brizard95,Cr2010,Cr2011,Cr2011a} still\ apply. Under the
circumstances, the very existence of kinetic equilibria remains dubious,
because of the possible absence of single-particle adiabatic invariants
which can survive in such a case.

Incidentally, it must be noted that in the customary non-relativistic
formulation of GKT, the possible inclusion of explicitly time-varying EM
fields is usually treated at most in asymptotic way and invoking
\textquotedblleft ad hoc\textquotedblright\ assumptions on their time
dependences \cite{dimit1}. Typical examples \cite{brizard95,hahm,PRL} are
provided by the treatment of small-amplitude and high-frequency
perturbations, having characteristic time-scales and scale-lengths
intermediate with respect to $\tau _{L}$ and $\Delta t$, and $r_{L}$ and $L$%
, with $r_{L}$ being the Larmor radius. However, in the relativistic context
space and time must be treated on equal footing, so that invariant scales
must be introduced preliminarily (see discussion below). Then, the issue
becomes whether suitably-large fluctuations of the EM 4-potential
(associated with the background EM fields) can be admitted on such scales.

The background to the present study is provided by the kinetic theory
recently established in Refs. \cite%
{Cr2010,Cr2011,Cr2011a,PRL,Cr2012,Cr2013,Cr2013b,Cr2013c,APJS} regarding
non-relativistic kinetic equilibria and their stability properties in the
case of collisionless magnetized plasmas subject to stationary or
quasi-stationary EM and gravitational fields. For this purpose, the method\
relying on the use of particle invariants was adopted. Based on the
identification of the relevant plasma regimes \cite{Cr2012}, the approach
allows for the implementation of a perturbative technique via suitable
expansions of the adiabatic invariants and, in turn, explicit
representations of the equilibrium kinetic distribution functions (KDFs)
holding in such cases. In particular, the conservation of the magnetic
moment was found to be related to temperature anisotropies in the
equilibrium KDF, producing in turn current flows and giving rise to kinetic
dynamo effects which lead to the self-generation of equilibrium magnetic
fields. The discovery of such a\ kinetic dynamo mechanism is relevant, as it
can in principle also operate in combination with other alternative
generation effects\ of different nature, whose existence has been recently
pointed out in Refs.\cite{Ase1,Ase2,Ase3}.

These conclusions were found to be ubiquitous, applying both to weakly or
strongly differentially-rotating axisymmetric astrophysical and laboratory
systems as well as spatially non-symmetric kinetic equilibria. In recent
developments, in addition, a further class of axisymmetric kinetic
equilibria has been pointed out which are independent of single-particle
energy \cite{PRE-new}. In these solutions the KDF is considered as a
function of the magnetic moment and the conserved canonical momentum only.
Remarkably, because of the absence of energy dependences, the latter type of
equilibria are absolutely stable with respect to infinitesimal axisymmetric
EM perturbations.

The goal of the present study is to investigate the possible extension of
these conclusions to relativistic plasmas characterized also by the
simultaneous presence of time-varying EM fields, in the sense specified
above. In detail, the scheme of the paper is as follows. Section 2 deals
with the construction of the extended gyrokinetic transformation which
permits to obtain a non-perturbative representation of the relativistic
particle magnetic moment. In Section 3 a covariant perturbative treatment is
developed in terms of a Larmor-radius expansion, permitting to obtain an
asymptotic representation of the magnetic moment and to display the
relationship between the relativistic and non-relativistic GKTs. In Section
4 the proof of existence of relativistic kinetic equilibria in axisymmetric
configurations is given, which are defined also in the presence of
time-varying EM fields. Section 5 then reports on the stability properties
of these relativistic equilibria, proving analytically the validity of
absolute stability criteria holding independent of the precise
representation of the equilibrium plasma distribution function. Finally,
concluding remarks are given in Section 6.

\section{Extended gyrokinetic transformation}

A prerequisite for the analysis reported here is to establish the adiabatic
conservation properties of the particle magnetic moment, when
radiation-reaction effects are ignored \cite{maha-2,EPJ5}. For this purpose,
a covariant formulation of gyrokinetics is adopted here, to describe a
plasma which is magnetized, in the sense that everywhere in the system $%
E^{2}-H^{2}<0$, with $E$ and $H$ denoting the eigenvalues of the Faraday
tensor. For definiteness, the background metric tensor $g_{\mu \nu }\left(
r\right) $ is considered prescribed, with $r$ denoting the dependence in
terms of the 4-position vector $r^{\mu }$.

The relativistic formulation of GKT then requires introducing an extended
gyrokinetic (GK) transformation of the 8-dimensional particle state $\mathbf{%
x}\equiv \left( r^{\mu },u^{\mu }\right) $, with $r^{\mu }$ and $u^{\mu
}\equiv \frac{dr^{\mu }}{ds}$ being the particle 4-position and 4-velocity,
and with $s$ being the particle proper-time. This is realized by a
diffeomorphism of the form \cite{Bek1,Bek}%
\begin{equation}
\mathbf{x}\equiv \left( r^{\mu },u^{\mu }\right) \longrightarrow \mathbf{z}%
^{\prime }\equiv \left( \mathbf{y}^{\prime },\phi ^{\prime }\right) ,
\label{87}
\end{equation}%
where $\phi ^{\prime }$ is a suitable gyrophase angle and $\mathbf{z}%
^{\prime }$ is constructed in such a way that its equations of motion are
gyrophase independent, namely $\frac{d}{ds}\mathbf{z}^{\prime }\equiv 
\mathbf{F}(\mathbf{y}^{\prime },s)$, where $\mathbf{F}$ is a suitable vector
field. In this regard, we notice that the theory developed in Refs.\cite%
{Bek1,Bek} can be re-formulated in such a way to apply in principle also in
a non-asymptotic sense.

Generalizing the non-relativistic approach developed in Ref.\cite{Cr2013},
this is achieved in terms of an extended guiding-center transformation of
the form%
\begin{eqnarray}
r^{\mu } &=&r^{\prime \mu }+\rho _{1}^{\prime \mu },  \label{1} \\
u^{\mu } &=&u^{\prime \mu }\oplus \nu _{1}^{\prime \mu },  \label{2}
\end{eqnarray}%
and then suitably-prescribing the vector $\mathbf{y}^{\prime }$ in terms of $%
\left( r^{\prime \mu },u^{\prime \mu }\right) $. Here the notation is
analogous to Refs.\cite{Bek1,Bek}. In particular, $r^{\prime \mu }$ is the
guiding-center position 4-vector, while primed quantities are all evaluated
at $r^{\prime \mu }$. Hence, $\rho _{1}^{\prime \mu }$ is referred to as the
relativistic Larmor 4-vector, while $\oplus $ denotes the relativistic
composition law which must warrant that $u^{\mu }$ is a 4-velocity. Notice
though that, on the rhs of Eq.(\ref{2}), while $u^{\prime \mu }$ is a
4-velocity, in principle the 4-vector $\nu _{1}^{\prime \mu }$ is not
necessarily required to be so.

We then introduce the orthogonal basis of unit 4-vectors $\left( a^{\mu
},b^{\mu },c^{\mu },d^{\mu }\right) $, with $a^{\mu }$\ and $\left( b^{\mu
},c^{\mu },d^{\mu }\right) $\ being respectively time-like and space-like
unit 4-vectors. By construction, the orientation of such a basis is
generally arbitrary and only depends on the particle 4-position, so that it
generally depends on 6 free parameters, corresponding to 3 pure space
rotations and 3 space-time rotations (boosts). A particular choice of
orientation is the one that associates the basis $\left( a^{\mu },b^{\mu
},c^{\mu },d^{\mu }\right) $\ to the antisymmetric Faraday tensor $F_{\mu
\nu }$, to be referred to in the following as EM-tetrad basis. Notice that
the EM-tetrad basis represents the natural covariant generalization of the
magnetic-related triad system formed by the orthogonal right-handed unit
3-vectors $\left( \mathbf{e}_{1},\mathbf{e}_{2},\mathbf{e}_{3}\equiv \mathbf{%
b}\right) $\textbf{\ }usually introduced in non-relativistic treatments \cite%
{Cr2013}. The existence of the EM tetrad relies on the fact that any
non-degenerate antisymmetric tensor $F_{\mu \nu }$\ has necessarily two
orthogonal invariant hyperplanes, which can be identified with the sets $%
\left( a^{\mu },b^{\mu }\right) $\ and $\left( c^{\mu },d^{\mu }\right) $.
One can show that each of the two invariant hyperplanes has a single
associated eigenvalue. In fact, assuming the signature of the metric tensor
to be $\left\langle 1+,3-\right\rangle $\ and denoting with $H$\ and $E$\
the 4-scalar eigenvalues of $F_{\mu \nu }$, one has%
\begin{eqnarray}
F_{\mu \nu }a^{\nu } &=&Eb_{\mu },\ \ \ F_{\mu \nu }b^{\nu }=Ea_{\mu }, \\
F_{\mu \nu }c^{\nu } &=&-Hd_{\mu },\ \ \ F_{\mu \nu }d^{\nu }=Hc_{\mu }.
\end{eqnarray}%
As a consequence of this formalism, in the EM-tetrad frame the Faraday
tensor can be represented as%
\begin{equation}
F_{\mu \nu }=H\left( c_{\nu }d_{\mu }-c_{\mu }d_{\nu }\right) +E\left(
b_{\mu }a_{\nu }-b_{\nu }a_{\mu }\right) .  \label{tetrad-fmunu}
\end{equation}%
Here it is assumed that both eigenvalues are non-vanishing. The physical
meaning is that $H$\ and $E$\ coincide with the observable magnetic and
electric field strengths in the reference frame where the electric and the
magnetic fields are parallel. The velocity transformation law relating the
EM-tetrad and the observer (laboratory) reference frame will be discussed
below. For the moment we remark that the EM-tetrad is constructed locally
and in such a way that locally the metric tensor $g_{\mu \nu }\left(
r\right) \cong \eta _{\mu \nu }$\ (condition of local flatness, in turn
based on the Einstein principle of equivalence), so that in the EM-tetrad
frame the controvariant and covariant basis $\left( a^{\mu },b^{\mu },c^{\mu
},d^{\mu }\right) $\ and $\left( a_{\mu },b_{\mu },c_{\mu },d_{\mu }\right) $%
\ are related locally by means of the Minkowski tensor $\eta _{\mu \nu }$.
More details on the definition of the EM tetrad reference frame can be found
in Refs.\cite{Bek1,Bek}.

When $u^{\prime \mu }$ is projected on the guiding-center EM-basis, it
determines the representation%
\begin{equation}
u^{\prime \mu }\equiv u_{0}^{\prime }a^{\prime \mu }+u_{\parallel }^{\prime
}b^{\prime \mu }+w^{\prime }\left[ c^{\prime \mu }\cos \phi ^{\prime
}+d^{\prime \mu }\sin \phi ^{\prime }\right] ,
\end{equation}%
where by construction%
\begin{equation}
u_{0}^{\prime }=\sqrt{1+u_{\parallel }^{\prime 2}+w^{\prime 2}}.  \label{u00}
\end{equation}%
Analogous decompositions follow also for $\rho _{1}^{\prime \mu }$ and $\nu
_{1}^{\prime \mu }$, which in general are expected to have non-vanishing
components along all the directions of the basis. Thus, denoting%
\begin{equation}
\left\langle h(\mathbf{z}^{\prime })\right\rangle _{\phi ^{\prime }}\equiv 
\frac{1}{2\pi }\oint d\phi ^{\prime }h(\mathbf{z}^{\prime })
\end{equation}%
as the gyrophase-averaging operator acting on a function $h(\mathbf{z}%
^{\prime })$, $\rho _{1}^{\prime \mu }$ and $\nu _{1}^{\prime \mu }$ are
assumed purely oscillatory by construction, so that $\left\langle \rho
_{1}^{\prime \mu }\right\rangle _{\phi ^{\prime }}=\left\langle \nu
_{1}^{\prime \mu }\right\rangle _{\phi ^{\prime }}=0$, while $r^{\prime \mu
} $, $u_{0}^{\prime }$, $u_{\parallel }^{\prime }$ and $w^{\prime }$ are
gyrophase-independent.

To obtain the particle phase-space extremal trajectory expressed via the GK
state $\mathbf{z}^{\prime }=\mathbf{z}^{\prime }\left( s\right) $, the
fundamental Lagrangian differential 1-form%
\begin{equation}
\delta \mathcal{L}\left( r,dr,u\right) =(u_{\mu }+qA_{\mu })dr^{\mu }
\end{equation}%
must be represented in terms of the same transformed state $\mathbf{z}%
^{\prime }$. Here $q\equiv \frac{Z_{j}e}{M_{0j}c^{2}}$ is the normalized
particle charge, with $M_{0j}$ and $Z_{j}e$ being the rest-mass and charge
of point-like particles belonging to the $j$th-species, while $A_{\mu
}\left( r\right) $ is the EM 4-potential associated with the external EM
field, which is assumed to be a smooth function of $r^{\mu }$. Dropping for
simplicity the particle species index $j$, the formal construction of GKT
then follows by adopting for $\delta \mathcal{L}$ a suitable gauge
representation \cite{Bek1}. Then, let us require that locally the
transformation (\ref{87}) is defined and the transformed differential form $%
\delta \mathcal{L}_{1}\left( z^{\prime },dz^{\prime }\right) $ is
gyrophase-independent, namely of the type%
\begin{equation}
\delta \mathcal{L}_{1}\left( y^{\prime },dy^{\prime },d\phi ^{\prime
}\right) \equiv \mathcal{L}_{1}\left( y^{\prime },\frac{dy^{\prime }}{ds},%
\frac{d\phi ^{\prime }}{ds}\right) ds,
\end{equation}%
with $\mathcal{L}_{1}$ being referred to as GK-Lagrangian.

As a consequence, the following non-perturbative representation is obtained
for the relativistic particle magnetic moment $m^{\prime }$:%
\begin{equation}
m^{\prime }=\left\langle \frac{\partial \rho _{1}^{\prime \mu }}{\partial
\phi ^{\prime }}\left[ \left( u_{\mu }^{\prime }\oplus \nu _{1\mu }^{\prime
}\right) +qA_{\mu }\right] \right\rangle _{\phi ^{\prime }}.  \label{m-exact}
\end{equation}%
The expression (\ref{m-exact}) is the covariant generalization of the
analogous non-relativistic result obtained in Ref.\cite{Cr2013}. A number of
important qualitative features must be pointed out. First, by construction $%
m^{\prime }$ is a 4-scalar, so that its value is frame-independent. Second,
consistent with non-relativistic theory, $m^{\prime }$ contains explicit
dependences in terms of all the independent components of the 4-velocity $%
u^{\prime \mu }$, namely $u_{\parallel }^{\prime }$ and $w^{\prime }$.
Finally, $A_{\mu }$ depends explicitly on the gyrophase via Eq.(\ref{1})
here.

\section{Perturbative theory}

In this section we determine a suitable covariant perturbative theory for
the analytical asymptotic treatment of the exact result obtained in the
previous section. In particular, a perturbative representation of $m^{\prime
}$ can be obtained \textquotedblleft a posteriori\textquotedblright\ by
Taylor-expanding $A_{\mu }\left( r\right) $ with respect to the Larmor
radius $\rho _{1}^{\prime \mu }$. Standard perturbative methods can be
adopted for this purpose \cite{Bek1,Bek,lit81,hahm}. The expansion is
performed invoking the following conditions:

1)$\ $The guiding-center 4-vector $\rho _{1}^{\prime \mu }$ is treated as an
infinitesimal, i.e. all its components are infinitesimal of $O\left(
\varepsilon \right) $ or are of higher-order, with $\varepsilon \ll 1$ being
a suitable scalar and dimensionless parameter. This is identified with the
invariant ratio%
\begin{equation}
\varepsilon \equiv \frac{r_{L}}{L}\ll 1,
\end{equation}%
where here%
\begin{equation}
r_{L}\equiv \sqrt{\rho _{1}^{\prime \mu }\rho _{1\mu }^{\prime }}.
\end{equation}%
In addition, $L$ is the invariant length-scale parameter defined as%
\begin{equation}
\frac{1}{L^{2}}\equiv \sup \left[ \frac{1}{\Lambda ^{2}}\partial _{\mu
}\Lambda \partial ^{\mu }\Lambda \right] ,
\end{equation}%
with $\Lambda \left( r\right) $ denoting\ either the set of scalar fields $%
E^{2}-H^{2}$ or the Kretschmer scalar $K\equiv R^{\alpha \beta \gamma \delta
}R_{\alpha \beta \gamma \delta }$ associated with the metric tensor. A
plasma characterized by the asymptotic ordering $\varepsilon \ll 1$\ will be
referred to as strongly-magnetized.

2)\ The 4-vector potential $A_{\mu }\equiv A_{\mu }\left( r\right) $\ and
the metric tensor $g_{\mu \nu }=g_{\mu \nu }(r)$\ are both analytic
functions which can exhibit fast space and time dependences, in the sense
that their Taylor expansions with respect to the \textquotedblleft Larmor
radius\textquotedblright\ $\rho _{1}^{\prime \mu }$, to be performed around $%
r^{\prime \mu }$ [see Eq.(\ref{1})], still converge asymptotically, in the
sense that everywhere%
\begin{eqnarray}
A_{\mu }\left( r^{\prime }\right) -A_{\mu }\left( r\right) &\sim
&O(\varepsilon )A_{\mu }\left( r\right) ,  \label{NON-UNIFORM-a} \\
g_{\mu \nu }\left( r^{\prime }\right) -g_{\mu \nu }\left( r\right) &\sim
&O(\varepsilon )g_{\mu \nu }\left( r\right) .  \label{NON-UNIFORM-b}
\end{eqnarray}%
Hence, both the 4-vector potential and the metric tensor vary on the same
scale $L$,\ so that they can be denoted as strongly non-uniform.

Thanks to these assumptions, upon carrying out the perturbative expansion to
first order in $\varepsilon $, it follows that Eq.(\ref{m-exact}) allows one
to determine the magnetic moment to the same accuracy, yielding $m^{\prime
}=\mu ^{\prime }\left[ 1+O\left( \varepsilon \right) \right] $, where%
\begin{equation}
\mu ^{\prime }=\frac{w^{\prime 2}}{2qH^{\prime }},
\end{equation}%
with $H^{\prime }$ denoting the corresponding guiding-center eigenvalue.
Similarly, one finds that to the leading-order, the relativistic Larmor
radius in the EM tetrad becomes $\rho _{1}^{\prime \mu }=r_{1}^{\prime \mu }%
\left[ 1+O\left( \varepsilon \right) \right] $, where%
\begin{equation}
r_{1}^{\prime \mu }=\frac{w^{\prime }}{qH^{\prime }}(c^{\prime }{}^{\mu
}\cos \phi ^{\prime }-d^{\prime \mu }\sin \phi ^{\prime }).
\end{equation}

Let us analyze the implications of these results. First, the magnetic moment
has been established to be an adiabatic invariant, which in principle can be
calculated by means of Eq.(\ref{m-exact}) to arbitrary order of accuracy in
terms of a power series expansion with respect to $\varepsilon $.
Manifestly, the\ present theory holds also in the non-relativistic limit in
which $u_{0}^{\prime }\cong 1$, $\left\vert u_{\parallel }^{\prime
}\right\vert ,\left\vert w^{\prime }\right\vert \ll 1$ and the space-time
can be treated as locally-flat on the scale $L$. Remarkably, due to the
arbitrariness in the choice of the functional dependences carried by the EM
4-potential, this means that both the electrostatic and vector potentials $%
\left( \Phi ,\mathbf{A}\right) $ are allowed in principle to have finite
space and time dependences, namely to be of the form $\Phi =\Phi \left( 
\mathbf{r},t\right) $ and $\mathbf{A}=\mathbf{A}\left( \mathbf{r},t\right) $.

The basic consequence is that GKT applies in all such conditions, so that
the magnetic moment is necessarily preserved as an adiabatic invariant under
much more general conditions than usually believed. The conclusion is that,
to retain properly explicit time-dependences of the EM field, a proper
formulation of the GK and guiding-center transformations requires generally
the adoption of a relativistic treatment involving extended phase-space
transformations of the type (\ref{87})-(\ref{2}).

The reason why a covariant formulation of GKT is necessarily required can be
easily understood. In fact, it must be remarked that the reference frame
that is at rest with the EM tetrad (EM tetrad frame) is the one in which the 
$\left( \mathbf{E}\times \mathbf{B}\right) $-drift velocity $\mathbf{V}_{E}$
vanishes by definition. In the case of a flat space-time, the latter is
given by%
\begin{equation}
\mathbf{V}_{E}=c\frac{\mathbf{W}}{2W^{2}}\left( 1-\sqrt{1-4W^{2}}\right) ,
\end{equation}%
where $\mathbf{W}\equiv \frac{\mathbf{E}\times \mathbf{B}}{B^{2}+E^{2}}$ 
\cite{LL}, and here $\mathbf{E}$ and $\mathbf{B}$ are the EM fields measured
in the laboratory reference frame. Therefore, if in such a frame 
\begin{equation}
\frac{\left\vert \mathbf{E}\times \mathbf{B}\right\vert }{B^{2}}\sim O\left(
1\right) ,  \label{ORD}
\end{equation}%
then\ $\mathbf{V}_{E}$ becomes relativistic, with $\mathbf{E}$\ and $\mathbf{%
B}$\ - and hence also $\mathbf{V}_{E}$ - still allowed to exhibit also fast
space and time dependences in the sense indicated above [see Eq.(\ref%
{NON-UNIFORM-a})]. In such a case, the appropriate GK treatment required for
single-particle dynamics is actually expected to be the relativistic one
adopted here. Indeed, when the previous orderings (\ref{NON-UNIFORM-a}) and (%
\ref{ORD}) apply: 1) the transformation between the laboratory and the EM
tetrad reference frames is necessarily a relativistic one; 2) $\mathbf{V}_{E}
$\ is strongly non-uniform. Analogous conclusions hold in the case of curved
space-time. As a result, in all cases particle dynamics may in general be
expected to be relativistic in the tetrad frame.

\section{Relativistic kinetic equilibria}

For the construction of kinetic equilibria in collisionless magnetized
plasmas, besides the magnetic moment $m^{\prime }$, the identification of
additional exact or adiabatic particle invariants is generally required. In
the case of non-stationary EM fields, it means that there is no coordinate
system in which the Faraday tensor admits an ignorable time-like coordinate.
Nevertheless, in such a case a possible occurrence is related to the
existence of spatial symmetries\ associated with the space-like coordinates
that have to be properly identified in a suitable reference frame. These
symmetries may characterize simultaneously both the Vlasov-Maxwell system
and the background metric tensor. For an illustration of the issue, we
consider here the case in which a symmetry of this type exists in an
arbitrary EM tetrad frame. Then the conjugate momentum is necessarily
conserved. This property is manifestly frame independent, so that it
actually must hold also in an arbitrary coordinate system different from the
previous EM-tetrad basis. In such a transformed reference frame the
corresponding conserved canonical momentum and particle magnetic moment must
be suitably related by means of the same coordinate-system transformation.

For the present treatment, it is therefore sufficient to require that the
space-like coordinate $Q$ is a symmetry coordinate in a prescribed EM
tetrad-frame. Then, the same ordering conditions introduced above for the
perturbative GK theory are invoked, dealing with the case of
strongly-magnetized plasmas. It follows that, neglecting corrections of $%
O\left( \varepsilon \right) $, the GK coordinate $Q^{\prime }\cong Q$ is
still a symmetry-coordinate for these plasmas. Thus, requiring $Q^{\prime }$
to be ignorable for the GK Lagrangian $\mathcal{L}_{1}$, the conjugate
canonical momentum%
\begin{equation}
P^{\prime }=\frac{\partial \mathcal{L}_{1}}{\partial \left( \frac{dQ^{\prime
}}{ds}\right) }
\end{equation}%
is necessarily conserved. A non perturbative representation of $P^{\prime }$
analogous to Eq.(\ref{m-exact}) can in principle be directly reached. Hence,
with respect to the EM tetrad frame, $P^{\prime }$\ is conserved provided $%
\frac{\partial \mathcal{L}_{1}}{\partial Q^{\prime }}=0$, namely both $%
g_{\mu \nu }^{\prime }$\ and $A_{\mu }^{\prime }$\ admit $Q^{\prime }$\ as
ignorable coordinate in such a frame. An example is provided by axisymmetry
with respect to the coordinate $Q^{\prime }=\varphi ^{\prime }$, in which
case one finds that to the leading-order in $\varepsilon $ the conjugate
canonical momentum $P^{\prime }=P_{\varphi ^{\prime }}$ is%
\begin{equation}
P_{\varphi ^{\prime }}=\frac{\partial r^{\prime \mu }}{\partial \varphi
^{\prime }}\left[ u_{0}^{\prime }a_{\mu }^{\prime }+u_{\parallel }^{\prime
}b_{\mu }^{\prime }+qA_{\mu }^{\prime }\right] ,  \label{p-fi}
\end{equation}%
with $u_{0}^{\prime }$ being given above by Eq.(\ref{u00}). The key element
shared by the two invariants $m^{\prime }$ and $P_{\varphi ^{\prime }}$ is
that they are expressed in terms of the same independent velocity-space
variables, namely $u_{\parallel }^{\prime }$ and $w^{\prime }$.
Nevertheless, explicit evaluation of $P_{\varphi ^{\prime }}$ requires the
preliminary construction of the EM tetrad (for details see Refs.\cite%
{Bek1,Bek}).

In the framework of the Vlasov-Maxwell statistical treatment the previous
results permit us to establish at once the general form of relativistic
kinetic equilibria. In fact, for the dynamical system associated with the
Lagrangian $\mathcal{L}$ and in the subset of phase-space in which $u^{\mu
}u_{\mu }=1$, the KDF\ $f=f\left( \mathbf{x}\right) $ is a 4-scalar which
obeys the covariant Vlasov equation%
\begin{equation}
\frac{df\left( \mathbf{x}\left( s\right) \right) }{ds}=u^{\mu }\frac{%
\partial f\left( \mathbf{x}\right) }{\partial r^{\mu }}+\frac{F^{\mu }}{M_{0}%
}\frac{\partial f\left( \mathbf{x}\right) }{\partial u^{\mu }}=0,
\label{vlavla}
\end{equation}%
where $F^{\mu }$ is the Lorentz force due to the external EM field.
Therefore, provided there are no other conserved quantities, a particular
solution is expressed as $f=f_{\ast }$, with $f_{\ast }$ being a
non-negative function of the GK particle invariants only, namely%
\begin{equation}
f_{\ast }=f_{\ast }\left( P_{\varphi ^{\prime }},m^{\prime }\right) .
\label{33}
\end{equation}%
A KDF that is a function of non-trivial particle invariants only is referred
to here as relativistic equilibrium KDF. Notice that, by construction, a KDF
of this type applies to strongly-magnetized plasmas for which $Q^{\prime }$\
is a symmetry coordinate in the EM tetrad and the magnetic moment $m^{\prime
}$\ can be treated according to the perturbative theory outlined above.

Inspection of Eq.(\ref{p-fi}) shows immediately that a necessary condition
for $f_{\ast }$ to be summable in velocity-space is that at least one of the
scalar products $\frac{\partial r^{\prime \mu }}{\partial \varphi ^{\prime }}%
a_{\mu }^{\prime }$ or $\frac{\partial r^{\prime \mu }}{\partial \varphi
^{\prime }}b_{\mu }^{\prime }$ never vanishes in configuration space.
Non-relativistic solutions of this type for stationary EM fields have been
referred to in Ref.\cite{PRE-new} as energy-independent kinetic equilibria.
Indeed, in the non-relativistic limit the scalar product $\frac{\partial
r^{\prime \mu }}{\partial \varphi ^{\prime }}a_{\mu }^{\prime }$ vanishes,
so that $P_{\varphi ^{\prime }}$ recovers its customary non-relativistic
expression (see Refs.\cite{Cr2011,Cr2011a}). In such a case a necessary
condition for summability is that the scalar product $\frac{\partial
r^{\prime \mu }}{\partial \varphi ^{\prime }}b_{\mu }^{\prime }$ must not
vanish (see also discussion in Ref.\cite{PRE-new}).

It is interesting to provide an example of relativistic equilibrium KDF of
the type (\ref{33}) expressed in terms of a generalized Gaussian
distribution \cite{Cr2010,Cr2011,Cr2011a,Cr2012,Cr2013,Cr2013b,Cr2013c,PRL},
namely%
\begin{equation}
f_{\ast }=\beta _{\ast }e^{-P_{\varphi ^{\prime }}^{2}\gamma _{\ast
}-P_{\varphi ^{\prime }}\delta _{\ast }-m^{\prime }\alpha _{\ast }},
\label{sol1}
\end{equation}%
where the set of functions $\left\{ \Lambda _{\ast }\right\} \equiv \left\{
\beta _{\ast },\gamma _{\ast },\delta _{\ast },\alpha _{\ast }\right\} $ are
referred to as structure functions, which can be either constant or
slowly-dependent on the same invariants, namely of the form $\Lambda _{\ast
}=\Lambda _{\ast }\left( \varepsilon ^{k}P_{\varphi ^{\prime }},\varepsilon
^{k}m^{\prime }\right) $ with $k\geq 1$. Notice that here the quadratic term
with respect to the canonical momentum $P_{\varphi ^{\prime }}$ warrants the
integrability of $f_{\ast }$ also in the non-relativistic approximation.

\section{Absolute stability criteria}

Following the approach pointed out in Ref.\cite{PRE-new} we now analyze the
stability of relativistic kinetic equilibria of the generic type (\ref{33}),
with $f_{\ast }$ to be assumed as analytic.\textbf{\ }For definiteness, we
consider here a strongly-magnetized plasma subject to infinitesimal
symmetric perturbations. By definitions, these are intended as perturbations
which do not exhibit a dependence on the (unperturbed) GK coordinate $%
Q^{\prime }$. Here it must be stressed in particular that the perturbations
of the KDF are constructed:\ 1) with respect to an unperturbed EM tetrad
frame; 2) in terms of the (related) unperturbed GK variables; and are
assumed to hold 3) in the subset of phase-space in which GK theory converges
asymptotically. As a consequence, in such a setting these perturbations
leave unchanged the same symmetry property assumed for the GK Lagrangian $%
\mathcal{L}_{1}$.

In principle, besides the KDF, these perturbations may involve both the EM
fields and the background metric tensor. Denoting these perturbations
respectively as $\delta f$, $\delta A^{\mu }$ and $\delta g_{\mu \nu }$, we
shall assume that they are characterized by an invariant length-scale $%
L_{osc}$ such that the dimensionless parameter $\lambda \equiv \frac{L_{osc}%
}{L}$ is ordered as%
\begin{equation}
\varepsilon \ll \lambda \ll 1.
\end{equation}%
Here $L_{osc}$ is defined as%
\begin{equation}
\frac{1}{L_{osc}^{2}}\equiv \sup \left[ \frac{1}{\delta \Lambda ^{2}}%
\partial _{\mu }\delta \Lambda \partial ^{\mu }\delta \Lambda \right] ,
\end{equation}%
with $\delta \Lambda \left( r\right) $ denoting the infinitesimal
perturbations of the scalar fields $\Lambda \left( r\right) $. Under these
conditions, both the canonical momentum $P^{\prime }$ and the magnetic
moment $m^{\prime }$ remain invariants by construction, although their
definition must be modified in accordance with the EM and gravitational
perturbations $\left( \delta A^{\mu },\delta g_{\mu \nu }\right) $ (see Ref.%
\cite{Bek1}).

It follows that $f_{\ast }$ remains an equilibrium KDF, while the
perturbation of the KDF $\delta f$ is of the general form 
\begin{equation}
\delta f=\delta f\left( \frac{y^{\mu }}{\lambda },y^{\mu },u^{\mu
},P^{\prime },m^{\prime }\right) .
\end{equation}%
Here $y^{\mu }$ denotes the position 4-vector which in the EM tetrad frame
can be treated, to the leading-order in $\varepsilon $ and neglecting
corrections of $O\left( \varepsilon \right) /O\left( \lambda \right) $, as
being independent of the ignorable GK coordinate $Q^{\prime }$. Notice that
necessarily $\delta f$ must satisfy the Vlasov equation in the Lagrangian
form 
\begin{equation}
\frac{d\delta f}{ds}=0.
\end{equation}

Under the condition of strongly-magnetized plasma, two cases can be
distinguished, as in Ref.\cite{PRE-new}. The first one corresponds to
consider only externally-produced perturbations due to non-vanishing $\left(
\delta A^{\mu },\delta g_{\mu \nu }\right) $ occurring at the initial
proper-time $s=s_{0}$ for which $\delta f\left( s_{0}\right) =0$. In such a
case the Vlasov equation requires identically that $\delta f\left( s\right)
=0$, which warrants stability of $f_{\ast }$. The second case is obtained by
assuming that also $\delta f\left( s_{0}\right) \neq 0$ (internally-produced
perturbations). Then, to the leading-order in $\lambda $ one obtains%
\begin{equation}
u^{\mu }\frac{\partial \delta f\left( \frac{y^{\mu }}{\lambda },y^{\mu
},u^{\mu },P^{\prime },m^{\prime }\right) }{\partial r^{\mu }}=0,
\end{equation}%
where the partial derivative is performed at constant $P^{\prime }$ and $%
m^{\prime }$. This means that $\delta f$ cannot exhibit fast dependences of
the type $\frac{r^{\mu }}{\lambda }$, so that it is actually of the form $%
\delta f=\widehat{\delta f}\left( y^{\mu },u^{\mu },P^{\prime },m^{\prime
}\right) $. To the next order in $\lambda $ and neglecting similarly
corrections of $O\left( \varepsilon \right) /O\left( \lambda \right) $, one
finds also that $\widehat{\delta f}$ must satisfy identically the equation (%
\ref{vlavla}), where again the partial derivatives are performed at constant 
$P^{\prime }$ and $m^{\prime }$. This implies that $\widehat{\delta f}$
itself must be a constant of motion. In view of the previous requirement on
the admissible conserved quantities, it follows that $\delta f$ is
necessarily an equilibrium solution of the form%
\begin{equation}
\delta f=\delta f\left( P^{\prime },m^{\prime }\right) .
\end{equation}%
Hence, it can always be absorbed in the definition of $f_{\ast }$, which is
therefore again stable.

These conclusions provide absolute stability criteria for relativistic
kinetic equilibria of the generic type (\ref{33}), in the sense that they
hold for arbitrary choices of the unperturbed KDF and of background EM and
gravitational fields satisfying the assumptions indicated above.

\section{Concluding remarks}

In summary, the notable implications of the GK theory developed in this
paper are that relativistic kinetic equilibria of the type (\ref{33}): 1)\
Exist provided the plasma is strongly-magnetized and is spatially-symmetric,
with the latter property being considered here as realized in the EM tetrad
frame. 2)\ Are consistent solutions of the relativistic Vlasov equation even
in the presence of strongly non-stationary and\textbf{\ }strongly spatially
non-uniform EM fields,\ together with a non-uniform background gravitational
field, all permitted to vary on the same invariant scale $L$. In this
regard, the functional form of the solution (\ref{33}) departs from that
considered in Ref.\cite{ko1} where energy-conserving systems were treated.
3)\ Are absolutely stable with respect to infinitesimal symmetric
perturbations which are either externally or internally produced.

It should be stressed that the present theory applies, of course, also in
the special case in which plasma particles appear as non-relativistic when
observed in the EM tetrad frame. As a result, the present formulation
includes as a particular case the\ theory recently discussed in Ref.\cite%
{PRE-new}.\textbf{\ }This means that the kinetic equilibria obtained here
generalize the energy-independent kinetic equilibria earlier investigated
[see Ref.\cite{PRE-new}] to a broader class of non-uniform EM fields. The
latter equilibria, in fact, were determined in the particular case of
stationary background EM fields.\textbf{\ }Based on the present formulation,
one can show that their validity can actually be extended to the case of
strongly non-stationary EM field [in the sense of Eq.(\ref{NON-UNIFORM-a})].%
\textbf{\ }The significant implication, besides the stability property
pointed out above, is that in the presence of time-varying EM fields which
are spatially-symmetric they represent the only possible class of
physically-admissible kinetic equilibria. This feature might have important
implications for the physical realization of these equilibrium
configurations in astrophysical and/or laboratory systems.

These results provide a convenient framework for the kinetic treatment of
both astrophysical and laboratory plasmas immersed in non-stationary EM
fields or intense radiation sources. The physical significance of this
achievement is of potential interest. In fact, plasma kinetic equilibria of
this kind persisting in radiation fields can also be expected to act as
transmitters of the radiation. This possibility, together with the
non-Maxwellian features which characterize intrinsically this kind of
energy-independent solutions (see discussion in Ref.\cite{PRE-new}), can be
relevant for the correct interpretation of observational data in
astrophysical scenarios where these conditions can be effectively realized,
for example as far as the thermal properties of these plasmas are concerned.

\textbf{Acknowledgments - }Financial support by the Italian Foundation
\textquotedblleft Angelo Della Riccia\textquotedblright\ (Firenze, Italy) is
acknowledged by C.C. Work developed within the research projects of the
Czech Science Foundation GA\v{C}R grant 14-07753P (C.C.)\ and GA\v{C}R
excellence grant 14-37086G (M.T. and Z.S.).

\end{document}